\documentclass[
  twoside,
  twocolumn,
  prl,
  preprintnumbers,
  floatfix,
  showpacs,
  superscriptaddress,
  footinbib,
]{revtex4}

\usepackage{amsmath}
\usepackage{amssymb}
\usepackage{color}
\usepackage{graphicx}
\usepackage{tabularx}
\usepackage{subfigure}
\usepackage{dcolumn}
\usepackage{times}

\newcommand{\RhN}[0]{\text{RhN}$_2$}
\newcommand{\RuN}[0]{\text{RuN}$_2$}
\newcommand{\PdN}[0]{\text{PdN}$_2$}
\newcommand{\OsN}[0]{\text{OsN}$_2$}
\newcommand{\IrN}[0]{\text{IrN}$_2$}
\newcommand{\PtN}[0]{\text{PtN}$_2$}

\renewcommand{\deg}{\ensuremath{^\circ}}

\newcommand{\fig}[1]{Fig.~\ref{#1}}

\newlength{\myhgt}

\begin{document}

\preprint{published as Phys. Rev. Lett. {\bf 100}, 095501 (2008) Copyright (2008) by the American Physical Society}
\title{
   Thermodynamic ground states of platinum metal nitrides
}

\author{Daniel \AA{}berg}
\email{aberg2@llnl.gov}
\affiliation{
  Lawrence Livermore National Laboratory,
  Chemistry, Materials and Life Sciences Directorate,
  7000 East Avenue, Livermore, California 94551, USA
}

\author{Babak Sadigh}
\email{sadigh1@llnl.gov}
\affiliation{
  Lawrence Livermore National Laboratory,
  Chemistry, Materials and Life Sciences Directorate,
  7000 East Avenue, Livermore, California 94551, USA
}
\author{Jonathan Crowhurst}
\affiliation{
  Lawrence Livermore National Laboratory,
  Chemistry, Materials and Life Sciences Directorate,
  7000 East Avenue, Livermore, California 94551, USA
}
\author{Alexander F. Goncharov}
\affiliation{
  Geophysical Laboratory,
  Carnegie Institution of Washington, 5251 Broad Branch Road
  NW, Washington, DC 20015, USA}

\begin{abstract}
  The thermodynamic stabilities of various phases of the nitrides of the platinum 
  metal elements are systematically studied using density functional theory. 
  It is shown that for the nitrides of Rh, Pd, Ir and Pt two new crystal structures, 
  in which the metal ions occupy simple tetragonal lattice sites, have lower formation 
  enthalpies at ambient conditions than any previously proposed structures. The region 
  of stability with respect to those structures extends to 17 GPa for \PtN{}. 
  Calculations show that the \PtN{} simple tetragonal structures at this pressure are 
  thermodynamically stable also with respect to phase separation. 
  The fact that the local density and generalized gradient 
  approximations predict different values of the absolute formation enthalpies as 
  well different relative stabilities between simple tetragonal and the pyrite or 
  marcasite structures are further discussed.


\end{abstract}


\maketitle

Until recently none of the Platinum-Metal (PM) elements (Pt, Ir, Os, Ru, Rh, Pd) 
were known to form stable stable compounds with nitrogen. Several such compounds 
have now been synthesized primarily under conditions of high static pressure and 
temperature (in the range of 60 GPa and 2000K respectively) \cite{soto}.
These include nitrides of Pt \cite{GreSan04,science},
Ir \cite{science,YouSan06}, Os \cite{YouSan06} and Pd \cite{CroGon07}
Except for \PdN{} all these compounds have been shown to be at least metastable
at ambient conditions.
The resulting crystal structures have been investigated by 
several groups both experimentally and theoretically \cite{GreSan04, science, YuZha05, CroGon07, YouSan06, YuZha06,YouMon06}, and by now consensus has been 
reached concerning the observed crystal structures and stoichiometry (one metal 
atom for every nitrogen dimer). \PtN{} and \PdN{} are formed in the 
pyrite crystal structure, a cubic phase with the metal atoms occupying fcc 
sites. The nitrogen dimers are centered around the fcc octahedral interstitial 
sites, oriented in all the four possible $\left<111\right>$ directions, such 
that all the nearest neighbor dimers make an angle of 70.53\deg\, with each 
other. A rotation of dimers such that two pairs point in the  
$\left[111\right]$ and  $\left[11\bar{1}\right]$ direction, respectively, 
results in the marcasite structure, the predicted ground state phase of 
\RuN{}, \RhN{} and \OsN{} \cite{YuZha06}. A small lattice distortion of 
marcasite then yields the monoclinic baddeleyite (CoSb$_2$) structure found 
in \IrN{}\cite{CroGon07,YuZha06}.   

The low-pressure phase diagrams of these systems are still not 
well understood. In particular their degree of thermodynamic stability has only
been addressed briefly \cite{science}, where it was proposed that the Pt-N 
compound is only metastable at low pressures. In this work, we address this 
issue by calculating the formation enthalpies of all the PM nitrides as a 
function of pressure. We also report the discovery of two new low-energy 
crystal structures that at low pressures are thermodynamically more stable 
than any of the crystal phases that have up to now been synthesized 
experimentally or calculated from theory. To our knowledge, these new 
crystalline phases have not previously been observed in any other compound. 

\begin{figure}[t!]
  \includegraphics[width=\columnwidth]{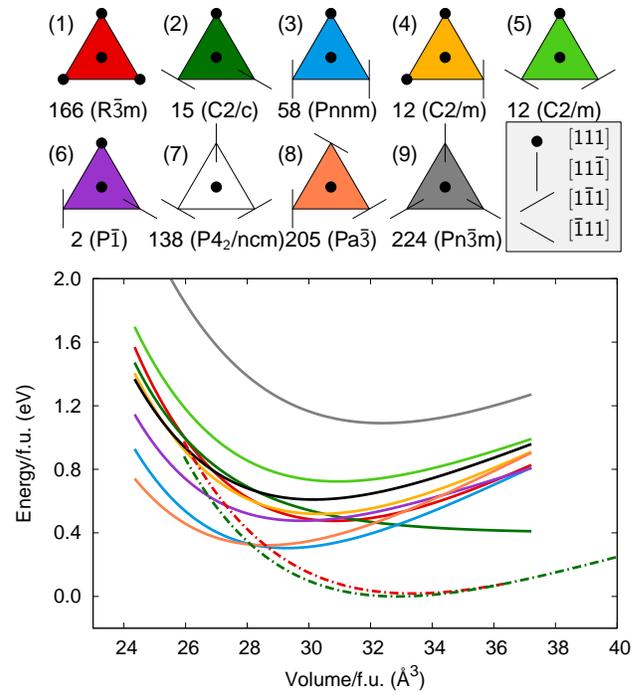}
  
  \caption{(Color online) Top: Projections of all possible distinct nitrogen 
    dimer orientations derived from the 12-atom conventional unit cell, along 
    with the space group number and Hermann-Maugin symbol of the corresponding 
    structure. Bottom: Energy/f.u. vs. volume curves from LDA for the above relaxed 
    structures (solid lines) in the case of \PdN{}. Each structure can be 
    identified by matching the line color to the corresponding triangle color. 
    Note that structure 7 corresponds to the black line. The green and red 
    dashed lines correspond to the ST$_\text{AA}$ and ST$_\text{AB}$ 
    structures, respectively.}
  \label{fig:projs}
\end{figure}

In spite of the apparent differences in the overall symmetries of the pyrite, 
marcasite and baddeleyite structures, they all belong to the same class
of crystal structures. They can be obtained by placing the PM atoms on an fcc
lattice and the center of $\left<111\right>$-oriented nitrogen dimers on the 
octahedral interstitial sites of this lattice and allowing for atomic 
relaxations. For each nitrogen dimer there are four possible orientations, 
and thus for 
crystals with the periodicity of the fcc unit cell having four distinct 
PM-N$_2$ units, there are nine different crystal structures, two of which are 
the pyrite and marcasite phases. Note that as mentioned above, the baddeleyite 
structure is merely a distorted marcasite structure. 

Geometrically each individual crystal structure in this class can be 
distinguished by the projections of its dimers centered at $0.5 a\hat x$, 
$0.5 a\hat y$, $0.5 a\hat z$ and $0.5 a\left(\hat x + \hat y + \hat z\right)$ 
onto a (111)-plane. The dimer projections and space groups of the resulting
crystal phases are displayed in the top panel of \fig{fig:projs}. For 
reference, structures (3) and (8) correspond to marcasite and pyrite, 
respectively. 
To assess their theoretical equation-of-state we have performed density 
functional theory calculations using projector augmented-waves as implemented in the 
Vienna ab-initio simulation package 
\cite{KreHaf93}.
Both atomic position and cell shape relaxations 
were performed. In fact, only structures (8) and (9) maintained cubic symmetry 
after relaxation, while the others showed deviations of various magnitude. 
The resulting LDA energy vs volume curves for \PdN{} are shown as the solid 
lines in the lower panel of \fig{fig:projs}. As expected, amongst these 
structures, pyrite and 
marcasite have the lowest energies in the low-pressure region. 

\begin{figure}
  \includegraphics[height=1.0in]{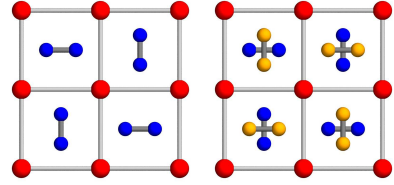}
  \caption{(Color online) Conventional unit cells of ST$_\text{AA}$ (left) 
    and ST$_\text{AB}$ (right) viewed along the $c$-axis. Red, blue and 
    yellow spheres represent metal ions, nitrogens in layer A and 
    nitrogens in layer B, respectively.
    \label{fig:ground}
  }
\end{figure}

Besides the nine structures mentioned above, we also show 
in Fig.~\ref{fig:projs} the corresponding curves for two additional structures 
(dashed lines). These are obtained by a continuous transformation from structure (2) 
by translating every other (001) plane in the $\left[100\right]$ direction 
(corresponding to a zone-boundary phonon mode). By allowing the lattice to 
relax by a tetragonal distortion in the $\left[001\right]$ direction, the metal
ions become arranged in a simple tetragonal (ST) lattice. During this 
transformation the nitrogen dimers that initially were oriented in the 
$\left<111\right>$ directions collapse into the (001) planes and now point 
in the $\left<100\right>$ and $\left<001\right>$ directions. 
In this way, a new class of crystal structures is created where 
metal nitride layers are stacked in different sequences. Each layer is
composed of metal ions occupying square lattice sites and 
face-centered nitrogen dimers pointing along the edges of the squares
and always being perpendicular to the four nearest-neighbor dimers. 
\fig{fig:ground} illustrates two particularly high-symmetry structures 
in this class with AA and AB stacking that have lower energies 
than any previously proposed crystal structure for most PM nitrides at ambient 
conditions. 
The first structure, hereafter denoted ST$_\text{AA}$,
is simple tetragonal and belongs to space group 127 (P4/mbm). The metal 
ions occupy the 2a Wyckoff positions and the nitrogen atoms are located at 
the 4g positions. The second structure, ST$_\text{AB}$, is monoclinic and belongs to 
space group 12 (P$2_{1}/$m). The metal ions occupy the 4e sites and the 
nitrogen atoms are located at the 4h and 4i positions. 

As is clearly seen in \fig{fig:projs}, the ST structures 
at low pressures are significantly lower in energy than both 
the pyrite and marcasite structures. Also, their equilibrium volumes 
are shifted towards larger volumes and and the bulk moduli are smaller.
Both LDA and GGA calculations yield small but finite densities-of-states 
at the Fermi level for the ST structures \cite{epaps}. However, it is well-known that these 
approximations often underestimate band gaps and e.g. predict Ge to be 
a metal. Hence, it is quite possible that the ST phases can be 
insulators with small band gaps. 

The high-pressure PM-nitride synthesis results from the past few years have 
established the extent of (meta-)stability of many of these 
compounds. Since they have not been observed in nature, it is reasonable to 
rule out that they are thermodynamically stable with respect to phase 
separation at ambient pressure and temperature conditions. Based on experiments
it is known that they are certainly stable at pressures above 50 GPa, and that 
\PtN{} (pyrite), \IrN{} (baddeleyite) and \OsN{} (marcasite), are at least 
metastable at zero pressure. 

There is however no real understanding either of the relative stabilities 
among the compounds or of their their degree of metastability at low-pressure 
conditions. These are very important issues for the synthesis and potential 
application of these materials. 
As it is yet difficult to assess their equilibrium phase diagrams 
experimentally we have performed extensive first principles calculations of 
the thermodynamic stabilities of the six PMN$_2$ compounds as a function of 
pressure.
 
The thermodynamic stability of any PMN$_2$ compound 
with respect to the separate phases as a function of pressure, is quantified
in terms of the formation enthalpy: 
\begin{align*}
  \Delta H &=  H_{\text{PMN}_2} - H_{\text {PM}} - H_{\text {N}_2}  
\end{align*}
\setlength{\myhgt}{3.4in}
\begin{figure*}
  \includegraphics[height=\myhgt]{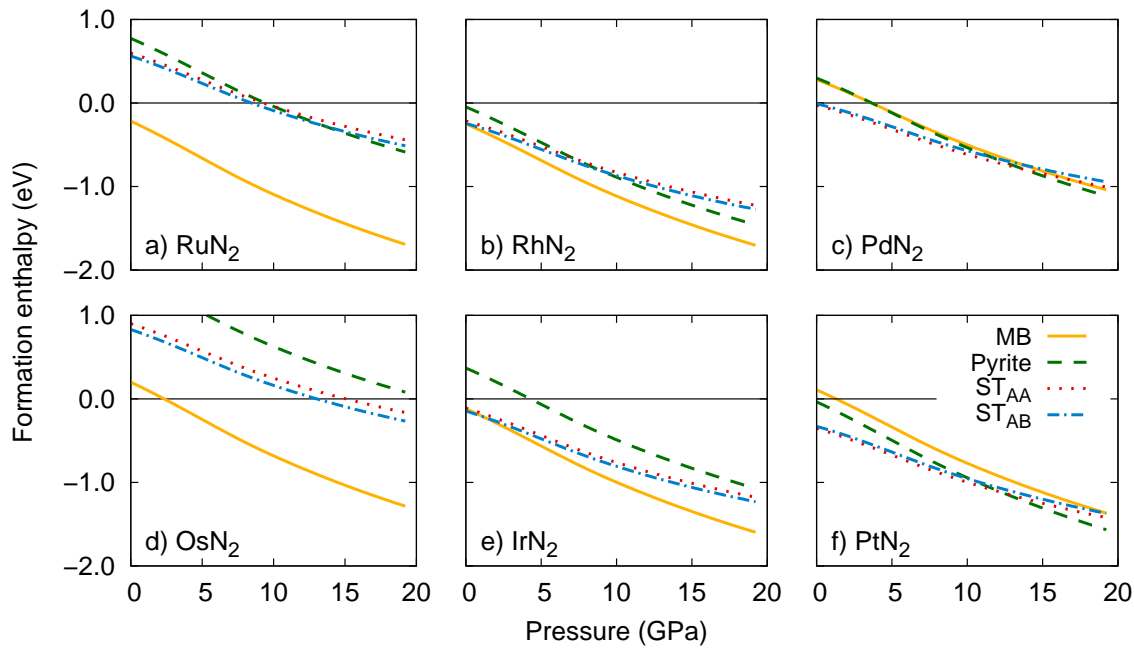}
  \caption{LDA Formation enthalpies per formula unit of a) \RuN{}, b) \RhN{}, c) \PdN{},
  d) \OsN{}, e) \IrN{} and f) \PtN{} as a function of pressure for marcasite 
  or baddeleyite (MB), pyrite and the two simple tetragonal structures.
  \label{fig:formLDA}
}
\end{figure*}

Under experimental conditions, nitrogen is 
molecular in its solid phase. 
At ambient pressure, it crystallizes 
into a hexagonal phase only at low temperatures 
and exhibits a polymorphic phase diagram undergoing several 
structural transformations upon increasing pressure. 
It is thus quite difficult to perform accurate free energy calculations 
for this system. 
Therefore, the most reliable way of estimating the 
enthalpy of this system as a function of pressure is to split it into 
contributions from the dimer energy and the cohesive energy:
  $H_{\text {N}_2} = 
  E_{\text {N}_2}^{\text{dimer}} + \Delta H_{\text {N}_2}^{\text{coh}}.$
We note that the dominant contribution to $H_{\text {N}_2}$
at ambient conditions is due to the bonding energy of the dimer, which 
can easily be calculated from first principles. We estimate the 
enthalpy of cohesion by calculating it
for the hexagonal phase at zero temperature and pressure 
within LDA/GGA and extract the change 
with pressure using the experimental equation-of-state data \cite{Oli90}.

\fig{fig:formLDA} shows the calculated formation enthalpies within the LDA.
For clarity, we consider here only the most important phases; 
pyrite, baddeleyite, marcasite and the two ST structures. As baddeleyite and 
marcasite always display similar energies, we have chosen to only include the 
structure with the lowest formation enthalpy.

We see that for the compounds formed with the group 
IX and X platinum metals, the ST structures become energetically favored 
at lower pressures, while the group VIII nitrides form in the 
marcasite/baddeleyite (MB) structure at all pressures. The range of the 
stability of the ST structures versus pyrite and MB is consistently pushed 
towards lower pressures for the group IX nitrides as compared with the 
group X nitrides. 

According to the LDA, the ST structures for \PdN{} and \PtN{} are stable up 
to 13 GPa, while for \IrN{} and \RhN{} they are only stable at very low pressures. 
The high pressure crystal structure of the group X nitrides 
is that of pyrite, while for all the earlier PM nitrides, the 
MB structures are the only structures observed at higher pressures. 

\begin{figure*}
  \includegraphics[height=\myhgt]{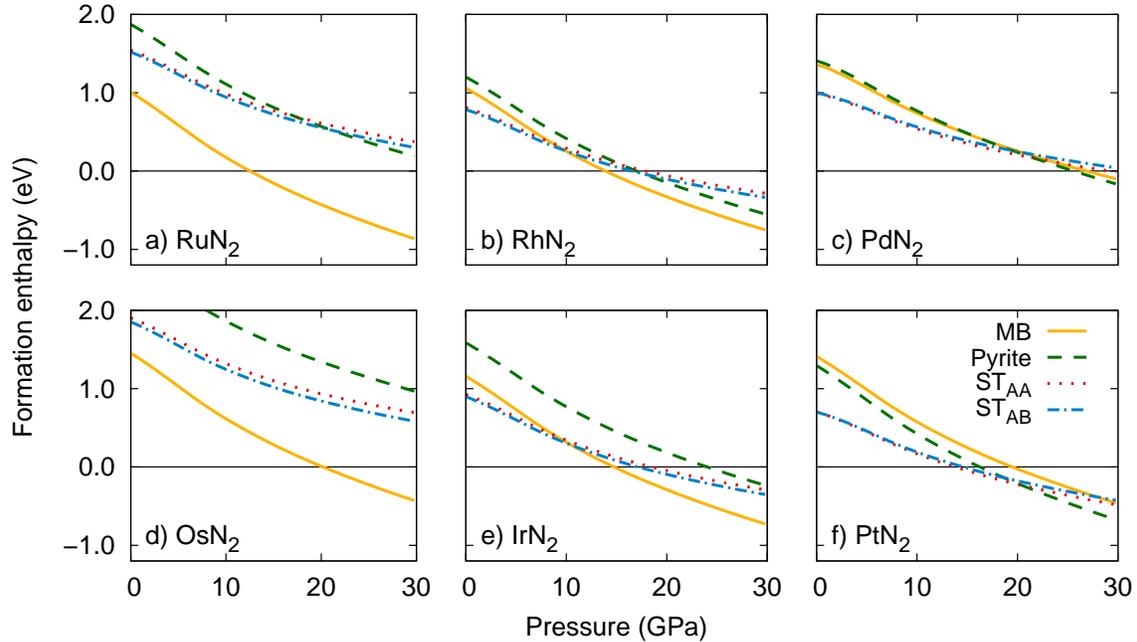}
  \caption{GGA Formation enthalpies per formula unit of a) \RuN{}, b) \RhN{}, c) \PdN{},
    d) \OsN{}, e) \IrN{} and f) \PtN{} as a function of pressure for marcasite 
    or baddeleyite (MB), pyrite and the two simple tetragonal structures.
    \label{fig:formGGA}
  }
\end{figure*}

The same trend is observed in the GGA-calculations, displayed in 
\fig{fig:formGGA}, except that the relative stabilities of the ST structures
are more pronounced as they constitute the low-enthalpy phases up to 20 GPa 
for \PdN{} and \PtN{} and up to 10 GPa for \RhN{} and \IrN{}.

A consistent feature observed in the phase diagrams is that the 
marcasite or baddeleyite structures are lowered in energy for the group VIII PM
nitrides and the pyrite and ST structures are lowered in energy for the group 
IX and X PM nitrides.

The formation enthalpies reported in \fig{fig:formLDA} also contain information
on the absolute thermodynamic stability of these compounds. According to the
LDA all but \OsN{} exhibit at least one stable phase with respect to the 
separate phases at ambient conditions. In fact, the lowest enthalpy phases 
for all considered PM-nitrides but \OsN{} are predicted to be stable at all 
pressures. This is in contrast to the results from the GGA, which predict
that the compounds are stable at 12 GPa for \RuN{}, 20 GPa for \OsN{}, around 
15 GPa for \RhN{}, \IrN{} and \PtN{} and 25 GPa for \PdN{}. 

In summary, we find from \fig{fig:formLDA} the remarkable result that according 
to the LDA two novel crystal structures, never observed before, 
constitute the ground state phases of the group IX and X PM nitrides. 
\PtN{} is the most stable among all these compounds, with a cohesive energy of 
about 0.3 eV/f.u. 
In contrast, GGA predicts all the PM nitrides to phase separate at ambient conditions. 
For example, the \PtN{} ST phases are unbound within GGA by 
0.6 eV/f.u. at zero pressure. They become thermodynamically stable at about 15 GPa. 
Upon further investigation of Figs. \ref{fig:formLDA} and \ref{fig:formGGA} we 
find that the difference between GGA and LDA enthalpies 
mainly stems from a rigid upward shift of the GGA curves by about 1.2 eV/f.u.  
This is not a surprising result as it is well known that LDA usually predicts larger 
cohesive energies and smaller lattice constants for solids than GGA. 
To compare with other nitrides, we find in the literature
LDA formation energies of zinc-blende GaN, InN, AlN compounds that are 
0.4 - 0.6 eV/f.u. lower than the corresponding values from GGA \cite{ZorBer01}. 
This is less than half the difference found for the PM nitrides in this work.
We attribute this to the presence of the nitrogen dimers.
To prove this point, we calculated the formation energy of PtN in 
the zinc-blende structure and found that the LDA prediction is smaller 
than the GGA by about 0.4 eV/f.u. 

Hence, the difference between the LDA and GGA 
cohesive energies of the PM nitrides is consistent 
with previous literature. As discussed above, 
this difference has important ramifications for the 
low-pressure phase stability of these compounds. Generally, GGA is expected to provide better
cohesive energies and equilibrium volumes than LDA. This is true for the 
zinc-blende nitrides (GaN, AlN, InN) mentioned above. However, 
in the case of PM nitrides, LDA seems to 
slightly outperform GGA with regards to lattice constants. 
For example, LDA predicts the lattice constant of \PtN{} to be  
4.80 \AA{}, in perfect agreement with experiment, as compared with the GGA lattice constant
of 4.88 \AA{} \cite{science}. Nevertheless, for the cohesive energies we believe that the GGA provides the most accurate values.

When comparing with experiments we find that the calculated
formation pressures are less than half the observed synthesis pressures. 
For example, 
the measured synthesis pressure for the nitrides of Pt and Ir is approximately 50 GPa 
\cite{GreSan04,science},
and 58 GPa for palladium nitride \cite{CroGon07}, while the calculated 
formation pressures within GGA for the same systems are 17, 15 and 27 GPa respectively.
At the experimental synthesis pressures, GGA predicts formation enthalpies of -1.4, -1.1 
and -1.4 eV/f.u., respectively for these systems.  
We attribute this discrepancy to the presence of large kinetic barriers to 
the formation of the PM nitrides when synthesized in the diamond anvil cell. 
The latter may be associated with for example a strain due to lattice 
mismatch between the compound and the parent metal on which it is grown.
The existence of large energy barriers to the formation of nitrides is well known in the 
synthesis literature. 
For example, according to GGA, GaN has a formation enthalpy of 
-1.7 eV/f.u. at zero pressure \cite{ZorBer01} but it requires a synthesis pressure of 2 GPa \cite{GrzJunBoc95}.

In conclusion, the LDA clearly overestimates the stability of the PM nitrides.
The GGA also predicts the transition pressures to be far below the experimental
values. These results are however consistent with the large kinetic barriers
found for zinc-blende nitrides. The only compound that exhibit a 
thermodynamically stable ST-phase is \PtN{}. The combination of large barriers 
to formation and small region of stability make the synthesis of these phases 
in the diamond anvil cell difficult. However, as we indicated earlier, these structures can be obtained
through martensitic transformations from marcasite or pyrite structures.  
Hence careful experimental work at pressures in the window of stability of 
of the ST structures may be 
required to produce these phases. Furthermore, since the associated Raman 
spectra are probably weak another primary in-situ diagnostic may be required 
such as x-ray diffraction.

This work was performed under the auspices of the U.S. Department of Energy 
by Lawrence Livermore National Laboratory under Contract DE-AC52-07NA27344.

\end{document}